%% file: TOP-14-003_temp.tex
\pdfoutput=1

\documentclass[11pt,twoside,a4paper,cmspaper,final,collab]{cms-tdr}
\def\svnVersion{338444}\def\cmsCernNoTag{CERN-PH-EP/2015-287}\def\cmsCernDate{\today}\def\cmsMessage{Published in the Journal of High Energy Physics as \href{http://dx.doi.org/10.1007/JHEP04(2016)035}{\doi{10.1007/JHEP04(2016)035.}}}

\begin{document}\cmsNoteHeader{TOP-14-003}

\hyphenation{had-ron-i-za-tion}
\hyphenation{cal-or-i-me-ter}
\hyphenation{de-vices}
\RCS$Revision: 330964 $
\RCS$HeadURL: svn+ssh://svn.cern.ch/reps/tdr2/papers/TOP-14-003/trunk/TOP-14-003.tex $
\RCS$Id: TOP-14-003.tex 330964 2016-03-07 07:52:48Z mojtaba $

\cmsNoteHeader{TOP-14-003}
\title{Search for anomalous single top quark production in
association with a photon in pp collisions at $\sqrt{s}=8$\TeV}

\date{\today}

\abstract{
The result of a search for flavor changing neutral
currents (FCNC) through single top
quark production in association with a photon is presented.
The study is based on proton-proton collisions at
a center-of-mass energy of 8\TeV using data collected with the CMS detector at the LHC,
corresponding to an integrated luminosity of 19.8\fbinv.
The search for $\PQt\gamma$ events where $\PQt\to\PW\PQb$ and
$\PW\to \mu\nu$ is conducted in final states with a muon, a
photon, at least one hadronic jet with at most one being consistent
with originating from a bottom quark, and missing transverse
momentum.
No evidence of single top quark production in association with a photon
through a FCNC is observed.
Upper limits at the 95\% confidence level are set on
the $\PQt\PQu\gamma$ and $\PQt\PQc\gamma$ anomalous couplings and translated into upper
limits on the branching fraction of the FCNC top quark decays:
$\mathcal{B}(\PQt\to\PQu\gamma) < 1.3\times 10^{-4}$
and $\mathcal{B}(\PQt\to\PQc\gamma) < 1.7\times 10^{-3}$.
Upper limits are also set on the cross section of associated $\PQt\gamma$
production in a restricted phase-space region. These are the most
stringent limits currently available.
}

\hypersetup{%
pdfauthor={CMS Collaboration},%
pdftitle={Search for anomalous single top quark production in
association with a photon in pp collisions at sqrt(s) = 8 TeV},%
pdfsubject={CMS},%
pdfkeywords={CMS, physics, top quark}}

\maketitle
\section{Introduction}

Evidence of physics beyond the standard model (SM) can be sought
in measurements of the rates of flavor changing neutral currents (FCNC) in
the top quark sector. Within the SM, top quark FCNC transitions are extremely suppressed by the GIM
mechanism~\cite{GIM}. The predicted branching fraction ($\mathcal{B}$) for
$\PQt\to\PQu\gamma$ and $\PQt\to\PQc\gamma$ decays are approximately $10^{-14}$~\cite{SM_tqa}.
However, an enhancement of several orders of magnitude is predicted in
some extensions of the SM, resulting in branching fractions observable at the LHC in some
cases~\cite{susy,tc}.
Therefore, observation of these rare top quark decay modes would be
indicative of physics beyond the SM.

Searches for FCNC $\PQt\PQu\gamma$ and $\PQt\PQc\gamma$ interactions have been
carried out by several experiments, with as yet no indication of a
signal. The measured upper limits at the 95\% confidence level (CL) on the branching fraction of
$\PQt\to\PQq\gamma$, with q representing an up or charm quark, through single top quark production are 4.1\% (L3)~\cite{l3}, 0.29\%
(ZEUS)~\cite{zeus1}, and 0.64\% (H1)~\cite{h1} . The 95\% CL limit set
by the CDF experiment through top quark pair production is $\mathcal{B}(\PQt\to
\PQq\gamma) < 3.2\%$~\cite{tevatron}.

The most general effective Lagrangian up to dimension-six operators, $\mathcal{L}_\text{eff}$,
used to describe the
FCNC tq$\gamma$ vertex has the following form~\cite{lagg}:
\begin{equation}\label{lag}
\mathcal{L}_\text{eff} =-eQ_{\PQt}\sum_{\PQq=\PQu,\PQc}\PAQq\frac{i\sigma^{\mu\nu} q_{\nu}}
{\Lambda}(\kappa^\mathrm{ L}_{\PQt\PQq\gamma}P_\mathrm{L}+\kappa^\mathrm{R}_{\PQt\PQq\gamma}P_\mathrm{R})\PQt A_{\mu}+\text{h.c.} ,
\end{equation}
where $e$ and $Q_{\PQt}$ are the electric charges of the electron and
top quark, respectively, $q_{\nu}$ is the four-momentum of the
photon, $\Lambda$ is an effective cutoff, which conventionally is
taken as the top quark mass, $\sigma^{\mu\nu} = \frac{1}{2}[\gamma^{\mu},\gamma^{\nu}]$,
and $P_\mathrm{L}$ and  $P_\mathrm{R}$ reflect, respectively, the left- and right-handed projection operators. The
strengths of the anomalous couplings are denoted by
$\kappa^\mathrm{L,R}_{\PQt\PQq\gamma}$.
No specific chirality is assumed  for the FCNC interaction of
$\PQt\PQq\gamma$, \ie, $\kappa^\mathrm{L}_{\PQt\PQq\gamma}=\kappa^\mathrm{R}_{\PQt\PQq\gamma}=\kappa_{\PQt\PQq\gamma}$.
In the SM, the values of $\kappa_{\PQt\PQu\gamma}$
and $\kappa_{\PQt\PQc\gamma}$ vanish at the lowest tree level.
A fully gauge-invariant effective-Lagrangian approach for parametrizing the top quark
FCNC interactions has been studied in Ref.~\cite{fabio}.
The FCNC effective Lagrangian can be used to calculate both
the branching fractions of the $\PQt\to\PQq\gamma$ decays and the
cross sections for the production of a
top quark in association with a photon.

The top quark FCNC processes can be probed through either
top quark production or decay. In this paper, we examine the associated
production of a single top quark and a photon, which is sensitive to the anomalous tq$\gamma$
FCNC coupling. The difference between quarks and antiquarks in the
parton distribution functions (PDF) of the proton in the presence of a
finite $\PQt\PQu\gamma$ coupling leads to an asymmetry between top and
anti-top quark production rates.  No asymmetry is expected for
 $\PQt\PQc\gamma$, because of the similar charm and anti-charm quark
contents in the proton. This would allow a distinction between the $\PQt\PQu\gamma$ and $\PQt\PQc\gamma$
signal scenarios if these processes were observed~\cite{mojtaba}.
Better sensitivity to the $\PQt\PQu\gamma$ coupling is expected because the
up quark PDF in the proton is larger than that of the charm quark.

Within the SM, top quarks can also be produced in association with a
photon. This proceeds through the radiation of a photon from the
initial- or final-state particles in $t$-channel, $s$-channel, and W-associated
production of single top quarks. These processes are treated as backgrounds
in this analysis.

\begin{figure}[htb]
\centering
\includegraphics[height=0.42\textwidth]{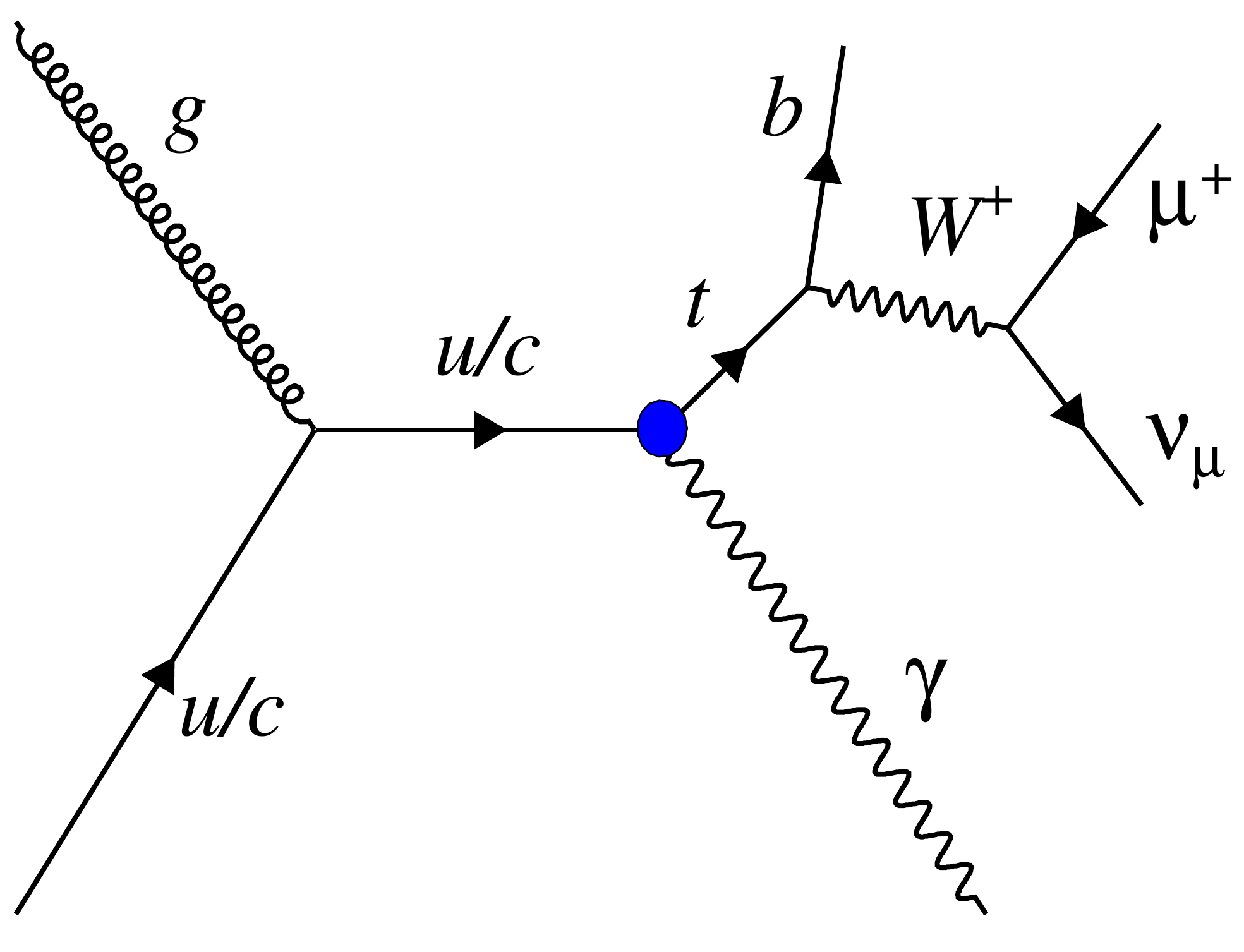}
\caption{Lowest-order Feynman diagram for single top quark production
  in association with a photon via a FCNC, including the muonic decay of the W boson from the top quark decay. The FCNC vertex is marked as a filled circle. }
\label{feynmann}

\end{figure}

We search for FCNC interactions at the $\PQt\PQu\gamma$ and $\PQt\PQc\gamma$
vertices by looking for events with a single top quark
and a photon in the final state, where the top quark  decays
into a W boson and a bottom quark, followed by the decay of the  W boson to
a muon and a neutrino. The final state includes $\PW^{\pm}\to \tau^{\pm} \nu_{\tau}$
events in which the $\tau$
lepton decays to $\mu\nu$.
 We focus on this particular leptonic decay
because it has a very clean signature.
Figure \ref{feynmann} illustrates the lowest-order diagram
for this $\PQt\gamma$ process
including the muonic decay of the W boson from the top quark decay.
The FCNC vertex is identified by a filled circle.

One of the distinctive signatures
of the signal is the presence of a high transverse momentum (\pt) photon in the
final state. The photon is expected to have large transverse momentum,
owing to its recoil from the heavy top quark.
The analysis is performed using events with a muon, a
photon, at least one hadronic jet, with at most one being consistent with
originating from a bottom quark, and missing transverse
momentum.
The results are compared with leading-order (LO)
and next-to-leading-order (NLO) calculations of the FCNC signal
production cross section based on perturbative quantum chromodynamics
(QCD)~\cite{nlosignal}.

\section{The CMS detector}

The central feature of the CMS apparatus is a superconducting solenoid
of 6\unit{m} internal diameter, providing a magnetic field of
3.8\unit{T}. A silicon
 pixel and strip tracker, a lead tungstate crystal electromagnetic
 calorimeter (ECAL), and a brass and scintillator hadron calorimeter
 (HCAL), each composed of a barrel and two endcap sections are contained within the superconducting solenoid volume.
 Extensive forward
 calorimetry complements the coverage provided by the barrel and endcap detectors.
Muons are measured in gas-ionization detectors embedded in the
 steel flux-return yoke outside the solenoid.

The first level of the trigger system, composed of custom hardware processors,
is designed to select the most interesting events in less than 4\unit{$\mu$s},
using information from the calorimeters and muon detectors.
The high-level trigger processor farm further decreases the event rate
from about 100\unit{kHz} to less than 1\unit{kHz}, before data storage.

A more detailed description of the CMS detector, together with a
definition of the coordinate
 system and kinematic variables used in this analysis, can be found in
 Ref.~\cite{cms}.

\section{Data and simulation samples}

The analysis is based on a data sample of proton-proton collisions at
a center-of-mass energy of 8\TeV, corresponding to an integrated
luminosity of 19.8\fbinv, collected with the CMS detector at the CERN LHC.

Monte Carlo (MC)  simulated signal samples of $\Pp\Pp\to\PQt\gamma\to\PW^{\pm}\PQb\gamma\to
\ell^{\pm}\nu_{\ell}\PQb\gamma$, with $\ell$ representing
\Pe, $\mu$, or $\tau$ leptons, are generated with the
PROTOS 2.0 generator~\cite{lagrangy}, with a minimum
\pt requirement of 30\GeV for the associated photon.
PROTOS is a LO generator for single top quark and \ttbar production
that includes anomalous top quark couplings.

To study the response of the analysis to the signal and to processes
with potentially  similar final-state signatures,
simulated event samples of  $\PQt+\gamma$, \ttbar, $\ttbar+\gamma$, $\PW\gamma+$jets,
$\Z\gamma+$jets, Drell--Yan, $\PW+$jets,
and $\PW\PW\gamma+\text{jets}$ events are generated using the LO \MADGRAPH 5
generator~\cite{madgraph}.
Diboson samples (WW, WZ, and ZZ) are generated using \PYTHIA 6
\cite{pythia}. Single top quark events from tq-, tb-, and tW-channel are generated with the NLO
\POWHEG 1.0
\cite{powheg,powheg2,powheg3,powheg4} event generator.
The NLO predictions for the main irreducible $\PW\gamma+\text{jets}$
background and the $\Z\gamma+\text{jets}$ process are calculated using the BAUR generator
\cite{baur}.

For all simulated samples, showering and hadronization are implemented
with \PYTHIA 6, and $\tau$ lepton decays with the \TAUOLA 2.7 program~\cite{tauola}.  The CTEQ6L~\cite{cteq6l} PDFs  are used to model the
proton PDFs for the LO generators, while CT10~\cite{ct10} is used for the NLO generators.
The top quark mass is set to 172.5\GeV.

The response of the CMS detector is simulated with \GEANTfour~\cite{geant}, and all simulated events are
reconstructed and analyzed using the standard CMS software. The MC
simulated events are weighted to reproduce  the trigger and
reconstruction efficiencies measured in data. The \PYTHIA 6 generator is used
to simulate the presence of additional proton-proton interactions
in the same or nearby proton bunch crossings (pileup).
The distribution of the number of pileup events in the simulation is weighted
to match that in data.

\section{Event selection and reconstruction of signal}
\label{kin}

The signal events are generally characterized by the presence of an
isolated energetic photon, a muon, significant missing transverse
momentum, and one b quark jet (b jet).
The presence of an isolated muon and an isolated photon provides a clean
signature for the signal. Events are initially selected with a
single-muon trigger, requiring a muon with a
minimum \pt of 24\GeV within the pseudorapidity
range $\abs{\eta}<2.1$.
 Events are also required to have at least  one well reconstructed pp interaction
vertex candidate~\cite{vertex}.
When more than one interaction vertex is found in an event, the
one with the highest $\sum \pt^{2}$ of its associated
charged-particle tracks is called the primary vertex and
selected for further analysis. The track associated with the muon candidate is required
to be consistent with a particle coming from the primary vertex.

A particle-flow algorithm (PF) is used to reconstruct single-particle candidates, combining information
from all subdetectors~\cite{pf,pf2}.
The muon candidates are reconstructed by matching the information for tracks in the silicon tracker and the muon system. The muon
candidates are required to have \pt$ > 26$\GeV and $\abs{\eta}<2.1$.
An accepted muon is required to have a relative
isolation $I_\text{rel} < 0.12$, where $I_\text{rel}$ is defined as the sum of the scalar \pt of all charged (except the muon candidate) and neutral PF
candidates inside a cone of size $\Delta R = \sqrt{\smash[b]{(\Delta \eta)^{2}+(\Delta\phi)^{2}}}< 0.4$
around the muon direction,
divided by the muon \pt, where $\Delta\eta$ and $\Delta\phi$ are the
differences in the pseudorapidity and azimuthal angle between the
directions of the PF candidate and the muon.
To remove the contribution from pileup,
the charged particles included in the calculation of $I_\text{rel}$ are
required to originate from the same vertex as the muon.
Based on the average deposited energy density of neutral particles
from pileup, a correction is applied to the neutral
component in the isolation cone.
One muon candidate is required in each
event, and events with additional muon candidates with \pt$> 10$\GeV, $\abs{\eta}<2.5$, and
 $I_\text{rel}<$ 0.2 are discarded.

Photon candidates with significant energy deposition in
the ECAL are required to have a $\pt > 50$\GeV, with $\abs{\eta}< 2.5$, but be outside of the transition region between the ECAL barrel and endcaps, $1.44<\abs{\eta}<1.56$.

The isolation of photon candidates is defined using
the following criteria: the ratio of the hadronic energy $H$ to the total
electromagnetic energy $E$ ($H/E$) inside a cone of size $\Delta R < 0.15$ around
the crystal containing the largest energy is required to be less than 0.05;
the second moment of the electromagnetic shower in $\eta$
($\sigma_{\eta \eta}$)~\cite{sigmaieta} is required to be less than
0.011\,(0.031) in the barrel (endcaps).
Separate charged- and neutral-hadron isolation criteria, defined as
the scalar sum of the \pt of all charged- or neutral-hadron PF candidates inside a cone
of size $\Delta R < 0.3$ around the photon
candidate, are applied. For the barrel, charged- and neutral-hadron
isolation values
are required to be less than 0.7\GeV and $0.4+0.04\,
\pt^{\gamma}$, while for the endcaps they are required to be less than 0.5\GeV
and $1.5+0.04\, \pt^{\gamma}$\GeV, respectively, where
$\pt^{\gamma}$ is the transverse momentum of the photon candidate.
The isolation criteria are corrected for additional
interactions in the same bunch crossing~\cite{rho}.
A pixel detector track veto is employed to minimize the misidentification of an electron as a photon.
Events with exactly one photon candidate are selected for further analysis.

Events with one or more electron candidates that pass loose selection
requirements of  $\pt >20$\GeV,  $\abs{\eta}<2.5$, and $I_\text{rel} < 0.15$ are rejected.
The electron $I_\text{rel}$ is defined in a manner similar to that for muons, using an isolation cone size
of $\Delta R < 0.3$.

Jets are clustered from the reconstructed PF candidates, using the infrared-
and collinear-safe anti-\kt algorithm with a distance parameter of
0.5~\cite{kt}.  The charged
hadrons originating from pileup interactions are excluded from
the clustered PF candidates, and the remaining contributions from neutral particles
are taken into account using a jet-area-based correction~\cite{rho}.
The momentum of a jet is defined as the vector sum of the
momenta of all particles in the jet, and
corrections to the jet energy are applied as a function of the jet \pt and $\eta$~\cite{jes}.
Only jets with $\pt>30$\GeV and $\abs{\eta}<2.5$ are considered in the analysis.

The combined secondary vertex (CSV) algorithm~\cite{btagpaper,csv} is used
 to identify jets originating from the hadronization of b quarks.  The
 algorithm combines the information from the secondary vertex and track impact parameters
into a likelihood discriminant, whose output distinguishes
between b jets and light-flavor jets.
The chosen cutoff on the value of the discriminant corresponds to a b tagging efficiency of
about 70\%, while the misidentification probability is $\approx$18\% for c jets,  and $\approx$1.5\% for  other jets~\cite{btagpaper,csv}.

To reduce the background from \ttbar and $\ttbar+\gamma$ processes, events
with  more than one identified b jet are rejected.
In events with no b-tagged jet, the jet with the largest value of the b tag discriminant is
chosen as the b jet candidate.
The missing transverse momentum vector, $\ptvecmiss$,
is defined as the negative vectorial sum of the momentum in the transverse plane of all PF objects.
Its magnitude, $\pt^\text{miss}$, is required to be greater than 30\GeV.
The direction of the photon candidate is required to be separated from
the directions of the muon and b jet candidates by $\Delta R(\mu,\gamma) > 0.7$ and
$\Delta R(\text{\PQb jet},\gamma) > 0.7$.

The top quark kinematic properties are reconstructed using the
muon and b jet four-momenta and $\ptvecmiss$. The \pt
of the undetected neutrino is assumed to be equal to the magnitude of $\ptvecmiss$,
while its longitudinal component is obtained by constraining the invariant
mass of the neutrino and muon to the world-average value of the W
boson mass~\cite{pdg}. When the resulting quadratic
equation has two real solutions, the one with the smaller absolute
value  of the longitudinal component of the neutrino momentum is taken~\cite{singletopcms}. When the solution is
complex, the real part is considered as the longitudinal $z$ component of the neutrino momentum.
The top quark candidate is reconstructed by combining the
reconstructed W boson and the b jet candidate.
Events with a reconstructed top quark invariant mass $m_{\mu\nu \PQb}$ within
130 to 220\GeV are selected for further analysis.
After all the selection criteria, signal efficiencies of 1.8\% and 2.4\%
are achieved from simulation for $\PQt\PQu\gamma$ and $\PQt\PQc\gamma$ signal events, respectively.

\section{Background estimation}
\label{bkg}

The main background contributions
arise from $\PW\gamma+$jets and $\PW+\text{jets}$ events, where the $\PW+\text{jets}$ background can mimic the signal when a jet is
misidentified as a photon.
The $\PW\gamma+$jets and $\PW+\text{jets}$ backgrounds are estimated from data, while
estimates for the backgrounds from single top quark
(tq-, tb-, and tW-channel), $\PQt+\gamma$, \ttbar, \ttbar+$\gamma$,
Z+$\gamma+$jets, Drell--Yan,  $\PW\PW\gamma+\text{jets}$, and diboson
backgrounds are calculated from the numbers of simulated events
passing the event selection, scaled to their theoretical cross
sections.

The contributions from the $\PW+\text{jets}$ and
$\PW\gamma+\text{jets}$ backgrounds are estimated from data using a neural
network (NN) discriminant formed from a combination of several
variables: the \pt of the photon and jet
candidates, the cosine of the angle between the momenta of the W boson and photon candidate,
the azimuthal angle between the momentum of the photon candidate and the missing transverse momentum, and
$H/E$. The NN is trained to distinguish these two sources of
background and its output is parametrized as:
\begin{equation}
F(x_\mathrm{NN}) = c_{\PW\mathrm{j}}S_{\PW\mathrm{j}}(x_\mathrm{NN})+
c_{\PW \gamma \mathrm{j}}S_{\PW \gamma \mathrm{j}}(x_\mathrm{NN})+bB(x_\mathrm{NN}),
\label{par}
\end{equation}
where $x_\mathrm{NN}$ is the neural network output,
$S_{\PW\mathrm{j}}(x_\mathrm{NN})$, $S_{\PW \gamma \mathrm{j}}(x_\mathrm{NN})$, and $B(x_\mathrm{NN})$
are, respectively, the normalized distributions for $\PW+\text{jets}$,
$\PW\gamma+\text{jets}$, and the sum of all other backgrounds, and $c_{\PW\mathrm{j}}$, $c_{\PW \gamma \mathrm{j}}$, and $b$ are the corresponding
fractions of each distribution. From previous limits, it is known that
any signal contribution will be small and is not included in
Eq.~\ref{par}. The effect of its possible presence is accounted for as a
systematic uncertainty.
The parametrization in Eq.~\ref{par}  is fit to the data, leaving the $\PW+\text{jets}$ and
$\PW\gamma+\text{jets}$ normalizations as free parameters.
Both the normalization and the distribution in the sum of all other
backgrounds, \ie,  the $b$ and $B(x_\mathrm{NN})$ terms, are obtained from simulation.
The distribution for $\PW+\text{jets}$, $S_{\PW\mathrm{j}}(x_\mathrm{NN})$, is obtained from data
 in a control region defined by requiring photons with wide electromagnetic showers
($\sigma_{\eta \eta}>0.011$ for the barrel and $\sigma_{\eta \eta}> 0.031$ for the endcap), and
no b-tagged jets, while keeping all other selection criteria the same as in the signal
region. The requirement of no b-tagged jets
ensures a high content of $\PW+\text{jets}$, suppressing thereby the
\ttbar and single top quark contribution. The distribution for $\PW\gamma+\text{jets}$, $S_{\PW \gamma\mathrm{j}}(x_\mathrm{NN})$,
is obtained from simulation.
The numbers of $\PW+\text{jets}$ and $\PW\gamma+\text{jets}$ events are determined from the fit
to the NN output distribution.

The fit results are taken as central values for the analysis,
and are assigned uncertainties that reflect the differences obtained when
varying the control region definition.
Additionally, an uncertainty is assigned accounting for the limited
knowledge of the contaminations from other SM backgrounds in the control sample, estimated
through a comparison with the results after subtracting
their expectations from simulation.
To take into account the uncertainties coming from the theoretical
predictions of the cross sections for the simulated backgrounds, the individual
cross sections are each varied
by $\pm$30\%~\cite{nb1,nb2,mcfmref} and the differences in the fitted
results with respect to the nominal fit are added in quadrature.

A total of 1794 events are selected
in data and, assuming no contribution from FCNC, $1805\pm 80$ events are expected, where the uncertainty is statistical.
The expected amount of SM background is dominated by the $\PW\gamma+\text{jets}$ process, amounting to 57\% of the total.
The contributions of $\PW+\text{jets}$, \ttbar, and $\Z\gamma+\text{jets}$ events are 16\%, 8\%, and
7\% of the total background events, respectively.
The remaining background events originate from t+$\gamma$,
\ttbar+$\gamma$, single top quark (tq+tb+tW),
W$\PW\gamma+\text{jets}$, and diboson production.

\section{Signal extraction}

Several discriminant variables are used to distinguish the signal
from the SM backgrounds. To achieve the best discriminating power, a multivariate
classification, based on boosted decision trees (BDT)~\cite{bdt,mva}, is used. One BDT is used for the $\PQt\PQu\gamma$ channel and
another for the $\PQt\PQc\gamma$ channel to take advantage of the
slight differences in their production.
For the $\PQt\PQu\gamma$ signal, the asymmetry between the top and
anti-top quark rates translates into a lepton charge asymmetry.
The lepton charge is therefore used as an input in training the BDT for the $\PQt\PQu\gamma$ signal.
Eight variables are chosen to construct the two BDTs. The BDT
input variables are: (i) \pt of the photon candidate,
(ii) b tagging discriminant, (iii) \pt of the b jet, (iv) \pt of the
muon  (only for $\PQt\PQc\gamma$), (v) $\cos(\vec{p}_{\mathrm{t}},\vec{p}_\gamma)$, the cosine of
the angle between the direction of the reconstructed top quark and photon, (vi)
$\Delta R(\text{b jet},\gamma)$, (vii) $\Delta R(\mu,\gamma)$, (viii) lepton charge (only for $\PQt\PQu\gamma$), and (ix)
jet multiplicity.

The \pt of the photon candidate is the
most important variable for separating
signal from background.
The \pt of the muon does not contribute significantly
to the discrimination of the $\PQt\PQu\gamma$ signal, and is therefore not used in
this case.
Each BDT is trained using simulated signal (either $\PQt\PQu\gamma$ or $\PQt\PQc\gamma$) and
$\PW\gamma+\text{jets}$, \ttbar, and diboson background events.
The distributions used as input to the BDT are obtained from data for
$\PW\gamma+\text{jets}$ and $\PW+\text{jets}$ and from simulation for the remaining background
contributions.  The $\PW+\text{jets}$ distributions are obtained from the same
control region as used for the NN inputs. Events with a reconstructed top quark mass in the
sideband region defined as $m_{\mu\nu\PQb} > 220$\GeV or $m_{\mu\nu\PQb} < 130$\GeV are used to obtain the $\PW\gamma+\text{jets}$ distributions.  The
sideband region is enriched in $\PW\gamma+\text{jets}$, with about 35\% contamination
from other background sources.  This contamination is subtracted using
an estimate from data for the $\PW+\text{jets}$ contribution and MC predictions for the remaining
background sources.

\begin{figure}[htb]
\centering
\includegraphics[height=0.31\textwidth]{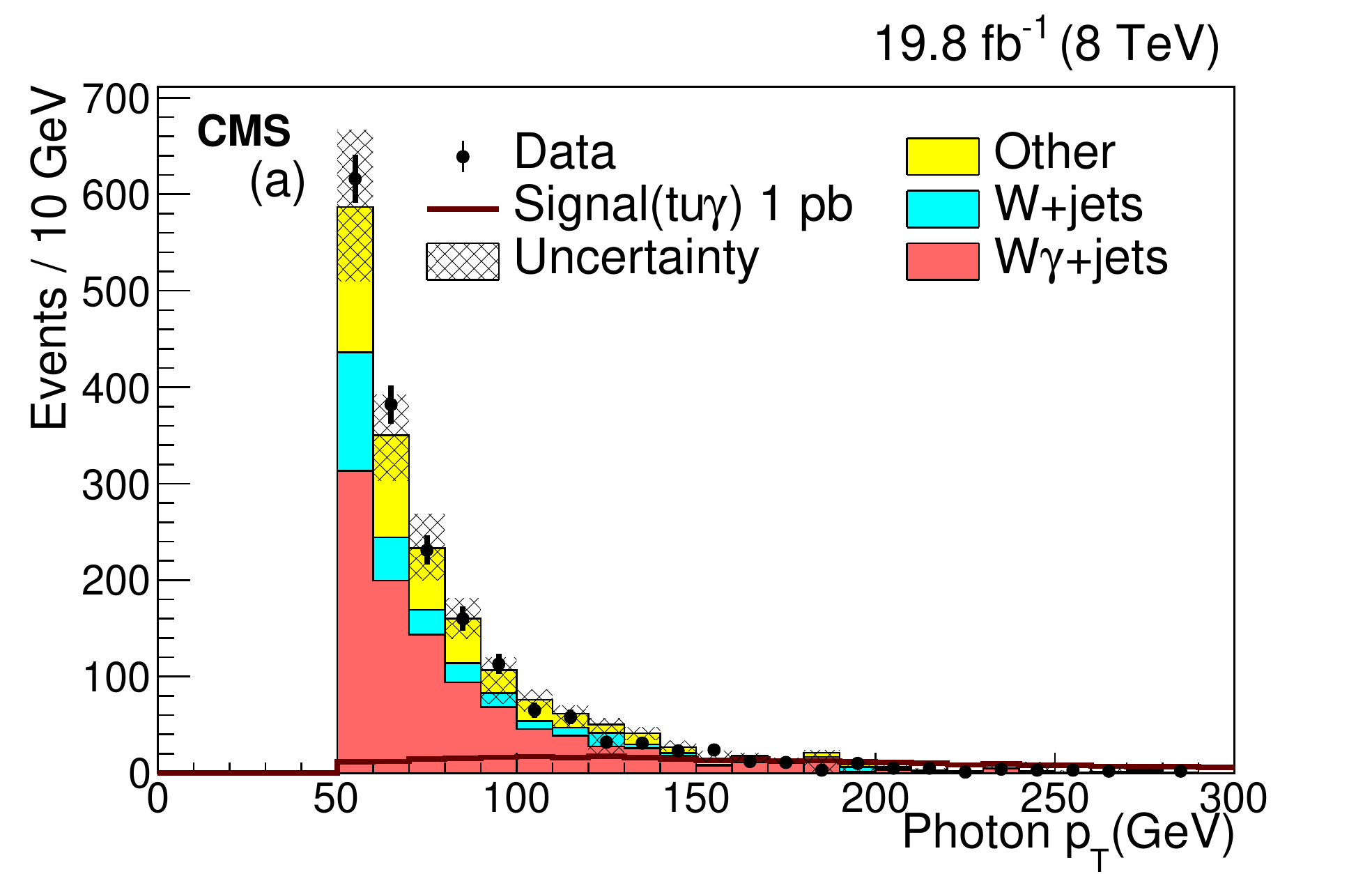}
\includegraphics[height=0.31\textwidth]{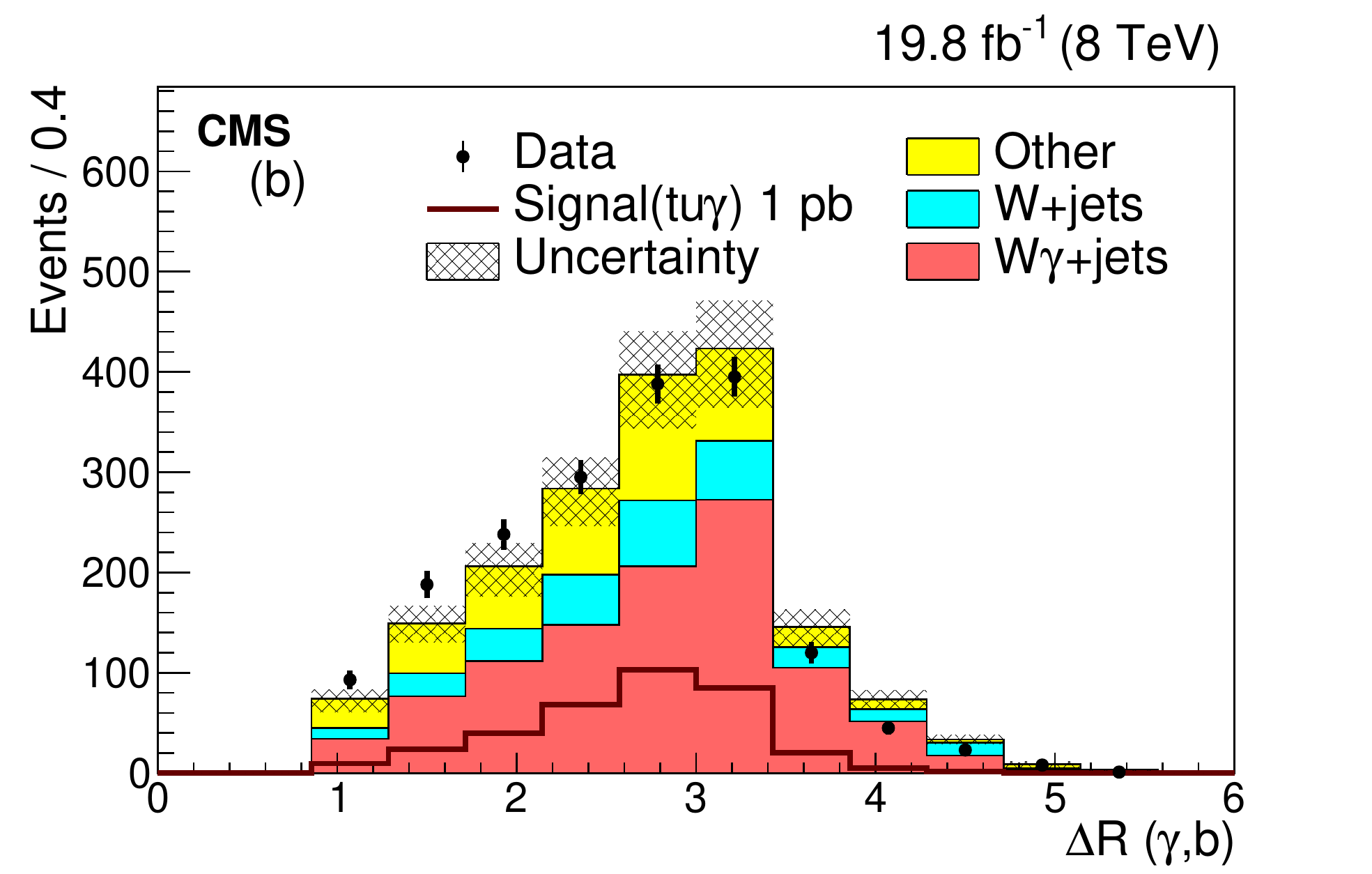}\\
\includegraphics[height=0.31\textwidth]{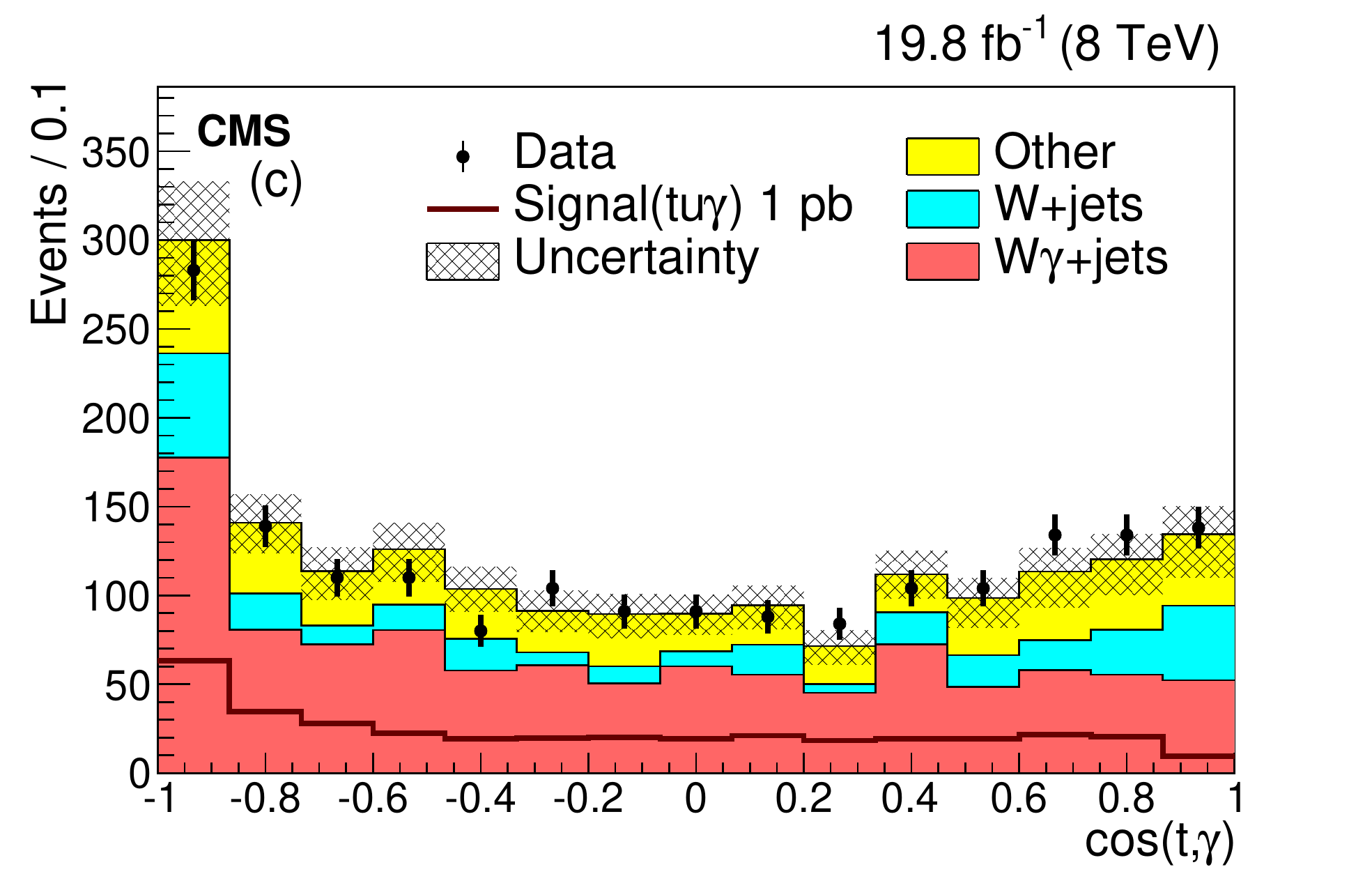}
\includegraphics[height=0.31\textwidth]{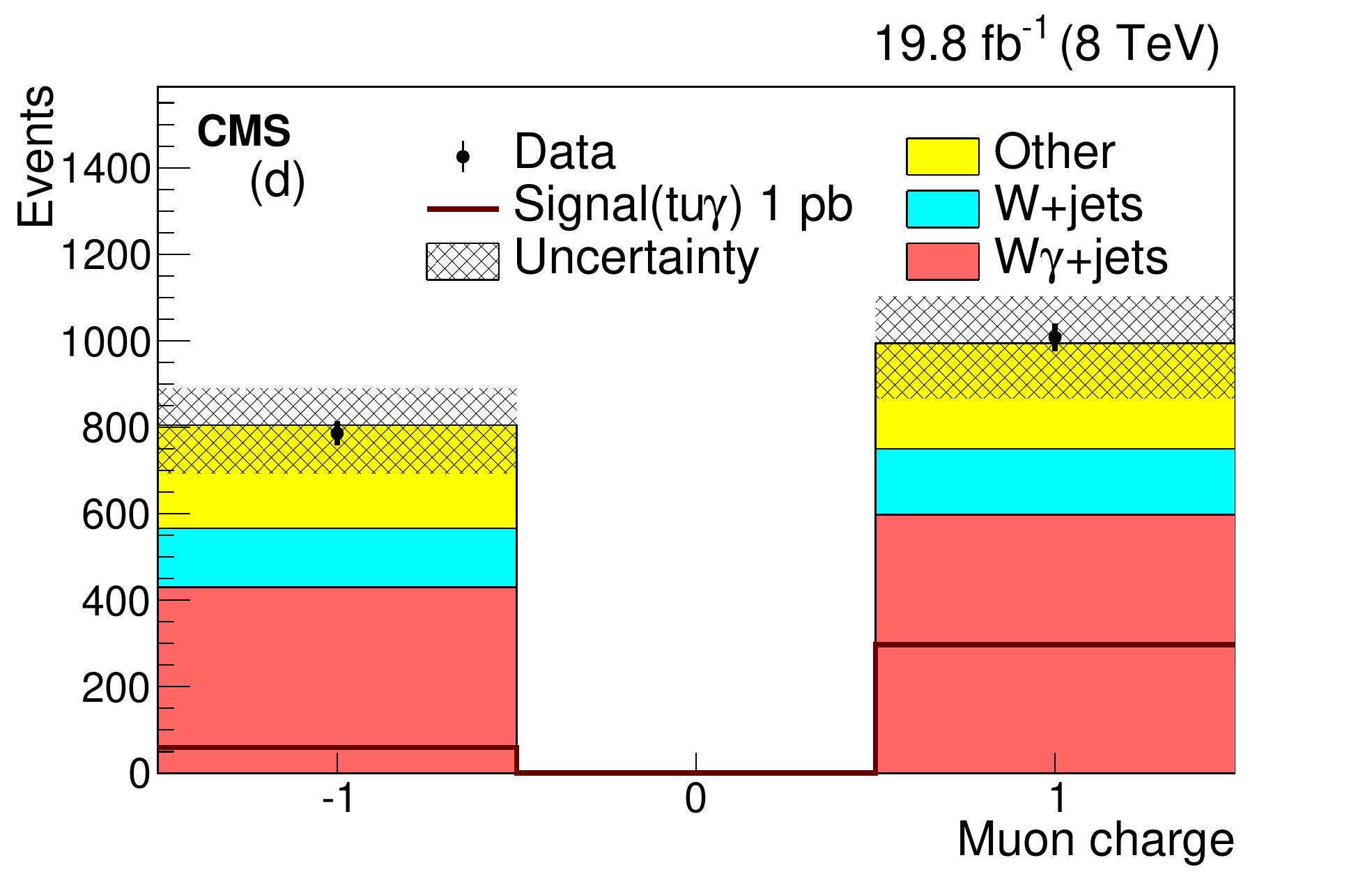}\\
\caption{Distributions of some of the input variables to the BDT:  (a) \pt of the
  photon, (b) $\Delta R(\gamma,\PQb)$,
(c) $\cos(\PQt,\gamma)$, and (d) muon charge
after the final event selection for data (points), the expected $\PQt\PQu\gamma$ signal (solid line), and background (histograms).
The $\PQt\PQu\gamma$
signal distributions are normalized to a cross section of 1\unit{pb}. The
vertical bars on the points show the statistical uncertainties in the data.
The hatched band shows the sum of the statistical and systematic
uncertainties in the estimated background combined in quadrature.
}
\label{photon}
\end{figure}

\begin{figure}[htb]
 \centering
\includegraphics[height=0.3\textwidth]{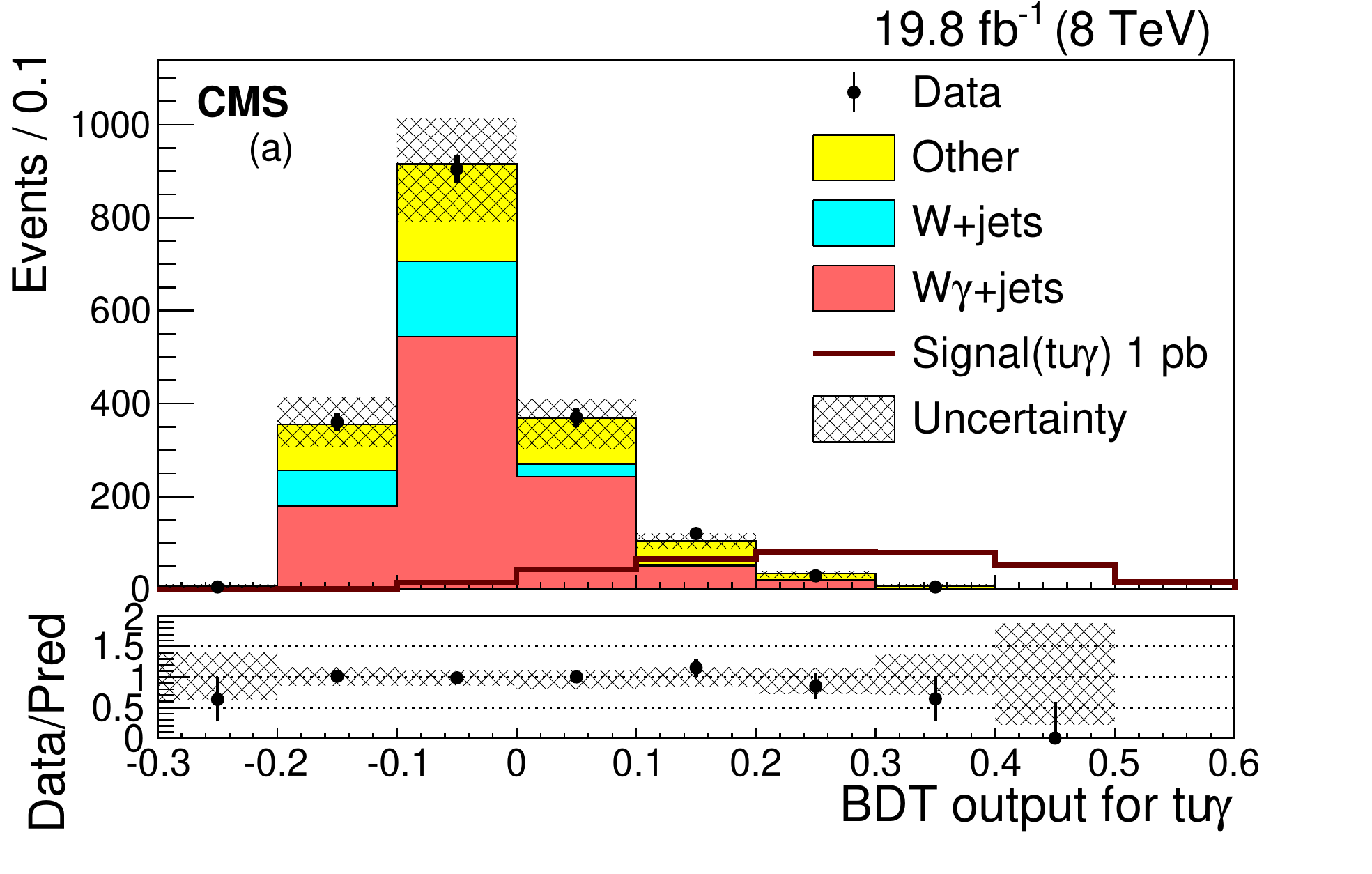}
\includegraphics[height=0.3\textwidth]{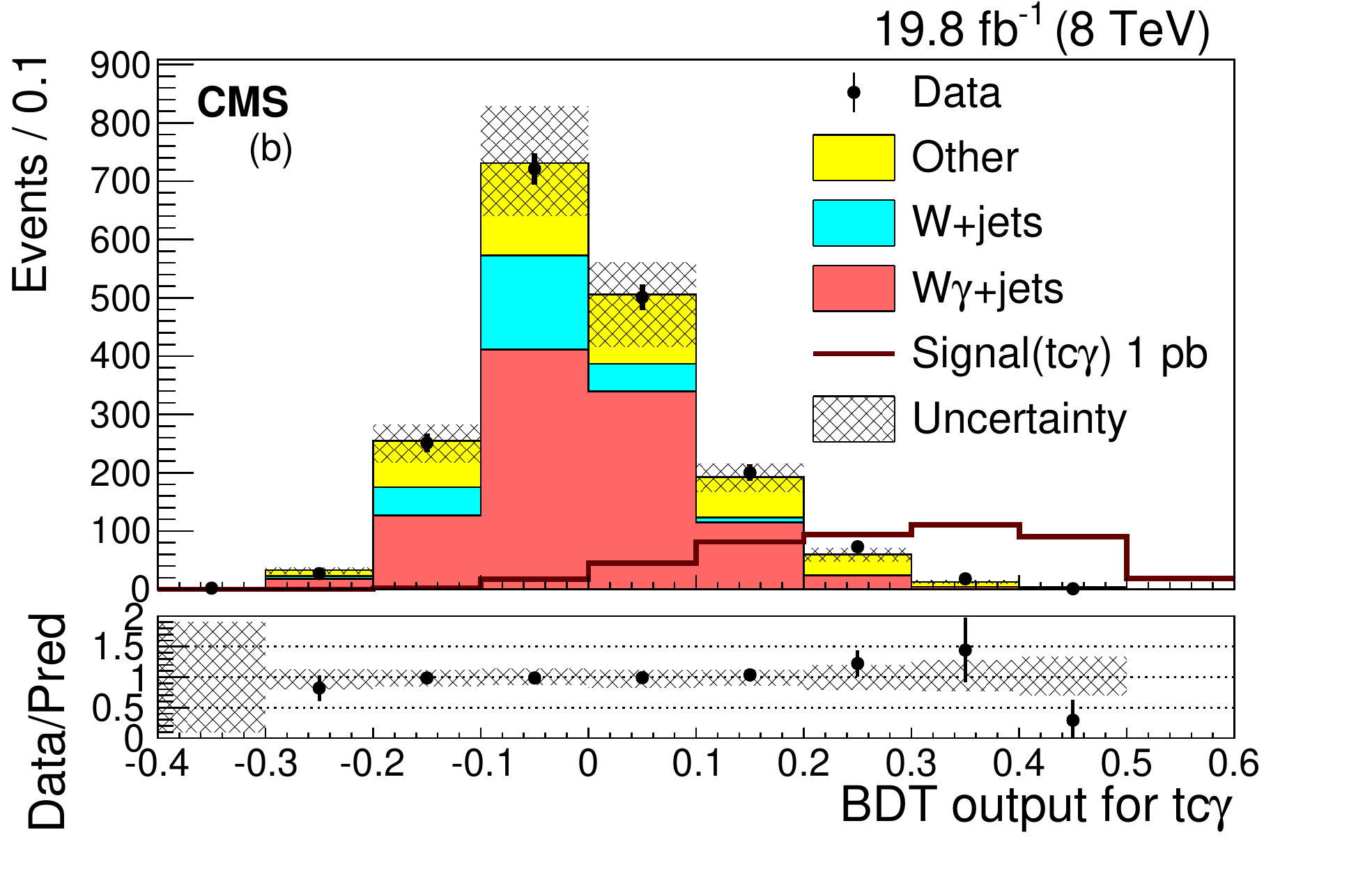}
    \caption{ The BDT output distributions for the data (points), the
      backgrounds (histograms),
       and the expected $\PQt\PQu\gamma$ (a) and $\PQt\PQc\gamma$ (b) signals (solid
       lines). The $\PQt\PQu\gamma$ and $\PQt\PQc\gamma$ signal distributions are
       normalized to a cross section of 1 pb. The vertical bars on the
       points give the statistical uncertainties. The hatched band
       shows the sum of the statistical and systematic uncertainties
       in the predicted background distributions combined in quadrature. The lower plots show the ratio of the data to the SM prediction.  }
       \label{output}
\end{figure}

Figure \ref{photon} shows the distributions of some of the BDT input variables for
the $\PQt\PQu\gamma$ signal and SM background.  Figure \ref{output} shows the BDT output
distributions for data, the estimated background, and the $\PQt\PQu\gamma$ and $\PQt\PQc\gamma$
signals.  As described above, the $\PW\gamma+\text{jets}$ and $\PW+\text{jets}$ distributions
and their normalizations are estimated from data, while the remaining
background contributions are obtained from simulation.  The signal
shapes are normalized to a cross section of 1\unit{pb} for showing the
expected signal distributions in the figures.  The vertical bars
indicate the statistical uncertainty.  The hatched band shows the
contribution of the statistical and systematic uncertainties added in
quadrature, with the dominant source being the statistical uncertainty
in the estimation of the number of  $\PW+\text{jets}$ and $\PW\gamma+\text{jets}$ events in data.

\section{Systematic uncertainties}
\label{syssec}

The effect on the signal and SM background expectations from different systematic
sources is discussed below.

\begin{description}
\item[Instrumental uncertainties:]
The uncertainties in the trigger efficiency~\cite{tge},
photon~\cite{phes} and lepton~\cite{muon} selection efficiencies, jet energy scale and resolution, missing
transverse momentum~\cite{jes}, and the modeling of pileup
are propagated to the uncertainties in the signal and SM background expectations.
The uncertainty in modeling the pileup is estimated by changing the
total inelastic proton-proton cross section by $\pm$5\%~\cite{pileup1}.
The uncertainty coming from the photon energy scale is estimated by changing the
photon energy in simulation by $\pm$1\% in the ECAL barrel and $\pm$3\% in the endcaps
\cite{phes}. The \pt- and $\eta$-dependent uncertainties in the b jet identification efficiencies
 and misidentification (mistag) rates are implemented as in Ref.~\cite{btagpaper}.
The systematic uncertainty in the measured integrated luminosity is
estimated to be 2.6\%~\cite{lumi}.
Among the instrumental uncertainties, the luminosity
uncertainty only affects the
normalization, while the uncertainties from the trigger, lepton and photon
selection efficiencies, b tagging, jet energy scale and resolution,
and pileup also affect the BDT discriminant output distributions for signal and background.
\item[Theoretical uncertainties:]
The uncertainty from the choice of PDF is determined according
 to the PDF4LHC prescription~\cite{pdf4lhc1,pdf4lhc2} using the MSTW2008~\cite{mstw} and
 NNPDF~\cite{nnpdf} PDFs.
The uncertainty from the factorization and renormalization scales is
evaluated by comparing simulated samples, produced using
factorization and renormalization scales multiplied and divided by a
factor of two relative to their standard values (top quark mass).
A conservative estimate of the uncertainty owing to the top quark mass
used in the simulation is obtained by producing simulated samples with
the top quark mass shifted by $\pm$2\GeV.
The uncertainties in the  PDF, renormalization and factorization scales,
and top quark mass affect both the predicted BDT distributions and the normalizations.
An uncertainty of 5\% in the signal rate is estimated from the NLO
QCD corrections~\cite{nlosignal}. This uncertainty is assumed
not to affect the signal distributions.
\item[Normalization of the background:]
The uncertainties described in Section~\ref{bkg} for the estimated
$\PW\gamma+\text{jets}$ and $\PW+\text{jets}$ backgrounds
are found to be  17\% and 23\%, respectively.
The uncertainties in the normalization of all other backgrounds
are found to be 30\%~\cite{nb1,nb2,mcfmref}.
\end{description}

\section{Upper limits on anomalous couplings}

No evidence is observed for anomalous single top quark production in association with
a photon in the BDT output distributions shown in Fig. \ref{output}.
These results are used to set an upper limit on this process,
as well as on the anomalous couplings $\kappa_{\PQt\PQu\gamma}$ and $\kappa_{\PQt\PQc\gamma}$.
The limits are  calculated using the modified frequentist approach~\cite{cls,cls2} that is implemented
in the \textsc{Theta} package~\cite{theta}. In this approach, a binned
maximum-likelihood
method is used for the BDT output distribution, which includes all systematic
uncertainties described in the previous section as nuisance parameters.
The NLO QCD corrections to the production of a single top quark plus a photon
through FCNC processes
are sizable and depend on the photon \pt requirement
\cite{nlosignal}.  Upper limits on the
cross sections are presented both with and without NLO QCD corrections.
We use a $k$ factor $k=\sigma_\mathrm{NLO}/\sigma_\mathrm{LO} = 1.375$
to go from LO to NLO, corresponding to a minimum photon \pt of 50\GeV~\cite{nlosignal}.

\begin{table*}[htb]
\centering
\topcaption{The expected and observed 95\% CL upper limits on
 the FCNC $\PQt\PQu\gamma$ and $\PQt\PQc\gamma$ cross sections times branching
 fraction $\mathcal{B}(\PQt\to\PW\PQb\to\PQb\ell\nu_{\ell})$, the anomalous couplings $\kappa_{\PQt\PQu\gamma}$ and
 $\kappa_{\PQt\PQc\gamma}$, and the corresponding branching
 fractions $\mathcal{B}(\PQt\to\PQu\gamma)$  and
 $\mathcal{B}(\PQt\to\PQc\gamma)$ at LO and NLO are given. The
 one and two standard deviation ($\sigma$)
ranges on the LO and NLO
expected limits are also presented.
\label{limittable}}
\begin{tabular}{ccccc}
\hline
         &  Exp. limit (LO)    &  ${\pm}1\sigma$ (exp. limit)  &
         ${\pm}2\sigma$ (exp. limit)    &   Obs. limit     (LO)  \\\hline
$\sigma_{\PQt\PQu\gamma}\,\mathcal{B}$ (fb) & 40
& 30--56  & 23--78 & 25 \\
$\sigma_{\PQt\PQc\gamma}\,\mathcal{B}$ (fb) &39
& 30--55  & 24--76 & 34  \\
 $\kappa_{\PQt\PQu\gamma}$ &0.036
&  0.032--0.043 & 0.028--0.051 & 0.029 \\
 $\kappa_{\PQt\PQc\gamma}$ &0.111
&  0.098--0.132 & 0.087--0.16 & 0.10 \\
 $\mathcal{B}(\PQt\to\PQu\gamma)$  & $2.7\times 10^{-4}$
&  $(2.0-3.8)\times 10^{-4}$ & $(1.6-5.4)\times 10^{-4}$ & $1.7\times
10^{-4}$  \\
 $\mathcal{B}(\PQt\to\PQc\gamma)$  & $2.5\times 10^{-3}$
&  $(1.9-3.6)\times 10^{-3}$ & $(1.5-4.9)\times 10^{-3}$ &
$2.2\times 10^{-3}$ \\ \hline
         &  Exp. limit (NLO)    &  ${\pm}1\sigma$ (exp. limit)   &
         ${\pm}2\sigma$
         (exp. limit)     &   Obs. limit     (NLO)  \\ \hline
$\sigma_{\PQt\PQu\gamma}\,\mathcal{B}$ (fb) & 39
&  30--58 & 25--84 & 26 \\
$\sigma_{\PQt\PQc\gamma}\,\mathcal{B}$ (fb) &42
& 29--59  & 22--86 & 37 \\
 $\kappa_{\PQt\PQu\gamma}$ &0.031
&  0.026--0.037 & 0.024--0.086 & 0.025 \\
 $\kappa_{\PQt\PQc\gamma}$ &0.098
&  0.082--0.12 & 0.071--0.140 & 0.091 \\
 $\mathcal{B}(\PQt\to\PQu\gamma)$  & $1.9\times 10^{-4}$
& $(1.4-2.9)\times 10^{-4}$  & $(1.2-4.2)\times 10^{-4}$ & $1.3\times 10^{-4}$ \\
 $\mathcal{B}(\PQt\to\PQc\gamma)$  & $2.0\times 10^{-3}$
&  $(1.3-2.7)\times 10^{-3}$ & $(1.0-4.0)\times 10^{-3}$ &
$1.7\times 10^{-3}$ \\ \hline
\end{tabular}
\end{table*}

\begin{figure}[htb]
  \centering
\includegraphics[height=0.41\textwidth]{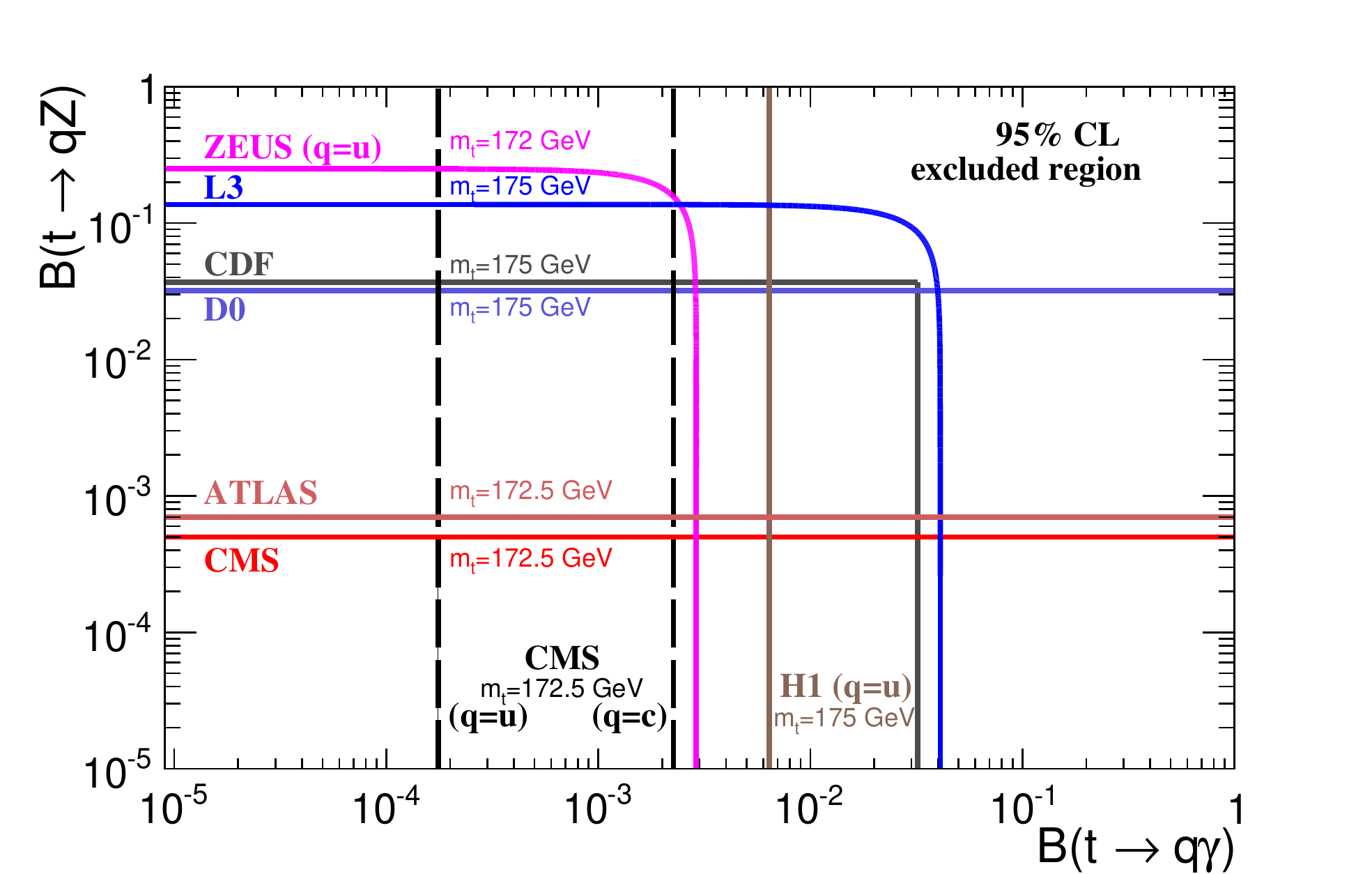}
    \caption{ The measured 95\% CL upper limits on $\mathcal{B}(\PQt\to\PQq\Z)$ versus $\mathcal{B}(\PQt\to\PQq\gamma)$
from the L3~\cite{l3}, ZEUS~\cite{zeus1}, H1~\cite{h1}, D0~\cite{d0z}, CDF
\cite{tevatron,cdfz}, ATLAS~\cite{atlasz}, and CMS experiments
\cite{cmsz}. The two
vertical dashed lines show the results of this analysis.}
       \label{comparison}
\end{figure}

The 95\% CL  upper limits on the number of events observed
are 9.1 and 16.0 for the $\PQt\PQu\gamma$ and $\PQt\PQc\gamma$ signals,
respectively.
The 95\% CL upper limits on the product of the LO signal cross
sections and the leptonic branching fraction
of the W boson are
$\sigma_{\PQt\PQu\gamma}\,\mathcal{B}(\PQt\to\PW\PQb\to\PQb\ell\nu_{\ell})<25$\unit{fb} and
$\sigma_{\PQt\PQc\gamma}\,\mathcal{B}(\PQt\to\PW\PQb\to\PQb\ell\nu_{\ell})<34$\unit{fb}. The corresponding upper
limits for the NLO calculations are
$\sigma_{\PQt\PQu\gamma}\,\mathcal{B}(\PQt\to\PW\PQb\to\PQb\ell\nu_{\ell})<26$\unit{fb} and
$\sigma_{\PQt\PQc\gamma}\,\mathcal{B}(\PQt\to\PW\PQb\to\PQb\ell\nu_{\ell})<37$\unit{fb}. The expected
limits and the one and two standard deviation limits on $\sigma_{\PQt\PQu\gamma}\, \mathcal{B}(\PQt\to\PW\PQb\to\PQb\ell\nu_{\ell})$
and $\sigma_{\PQt\PQc\gamma}\,\mathcal{B}(\PQt\to\PW\PQb\to\PQb\ell\nu_{\ell})$ at LO and NLO are presented in Table \ref{limittable}.
These results can be translated into upper limits
on the anomalous couplings $\kappa_{\PQt\PQu\gamma}$ and
$\kappa_{\PQt\PQc\gamma}$ and on the branching fractions $\mathcal{B}(\PQt\to\PQu+\gamma)$ and $\mathcal{B}(\PQt\to\PQc+\gamma)$  using
the theoretical expectations~\cite{saavi}.
The 95\% CL upper bounds on the anomalous couplings and branching fractions with and without
including the NLO QCD corrections to the signal cross
section are presented in Table \ref{limittable}, along with the
expected limits.
The one and two standard deviation ranges of the LO and NLO expected
limits on the anomalous
couplings and branching fractions are also shown in Table \ref{limittable}.
The measured 95\% CL upper limits on $\mathcal{B}(\PQt\to\PQq\Z)$ versus $\mathcal{B}(\PQt\to\PQq\gamma)$
from the L3~\cite{l3}, ZEUS~\cite{zeus1}, H1~\cite{h1}, D0
\cite{d0z}, CDF~\cite{cdfz}, ATLAS~\cite{atlasz}, and CMS~\cite{cmsz}
experiments, as well as the results of this analysis,
are presented in Fig. \ref{comparison}.

Table \ref{systab} summarizes the sources of the systematic uncertainties
in the expected upper limits on the signal cross sections. These are calculated as the ratio of the
difference of the shifted expected limit coming from the related systematic source and the nominal
expected limit.

\begin{table}[thb]
\centering
\topcaption{The sources and values of systematic uncertainties used to determine the
observed and expected upper limits on the $\PQt\PQu\gamma$ and $\PQt\PQc\gamma$ cross sections.
The values are given as a percentage of the expected upper limits. The sources are broken up into those that
only affect the overall rate of signal events and those that
affect both the rate and the shape of the BDT distributions.}
\label{systab}
\begin{tabular}{clcc}
\hline
Type       & \multicolumn{1}{c}{Source}
& $\PQt\PQu\gamma$ (\%)
& $\PQt\PQc\gamma$ (\%)                                                                                                                            \\ \hline
Rate       & \begin{tabular}[c]{@{}l@{}}Integrated luminosity\\ Background normalization ($\PW+\text{jets}$)\\ Background normalization ($\PW\gamma+\text{jets}$)\\ Other background normalizations\end{tabular}                                                                              & \begin{tabular}[c]{@{}c@{}}$1.8$\\ $5.6$\\ $2.5$\\ $<$1\end{tabular}                                                                  & \begin{tabular}[c]{@{}c@{}}$4$\\ $3$\\ $1.1$\\ $1$\end{tabular}                                                                        \\ \hline
Rate+Shape & \begin{tabular}[c]{@{}l@{}}Trigger efficiency\\ Pileup
  effects\\ Lepton identification and isolation\\ Photon
  identification and isolation\\ Photon energy scale\\ b tagging and mistag efficiency\\ Jet energy scale\\ Jet energy resolution\\ PDF \\ Scale\\ Top quark mass\end{tabular} & \begin{tabular}[c]{@{}c@{}}$2.2$\\ $7$\\ $<$1\\ $1.9$\\ $<$1\\ $1.1$\\ $2.9$\\ $2.1$\\ $3.1$\\ $1$\\ $2.5$\end{tabular} & \begin{tabular}[c]{@{}c@{}}$0.4$\\ $2.3$\\ $4.4$\\ $4.5$\\ $3.1$\\ $4$\\ $2.2$\\ $3.4$\\ $<$1\\ $2.4$\\ $1$\end{tabular} \\ \hline
\end{tabular}
\end{table}

\section{Upper limits on the FCNC cross sections for a restricted phase space}

Upper limits on the signal cross sections are also determined for a
restricted phase-space region in which
the detector is fully efficient. This removes the need to extrapolate
to phase-space regions where the analysis has little or no sensitivity.
The results are especially useful for comparing with theoretical
models that predict enhancements in
a particular phase-space region~\cite{fabio}.

The measurement uses a simpler event-counting procedure instead of a
fit to the BDT distribution.
 We define the fiducial
cross section, $\sigma_\text{fid}$, in a volume defined for  stable particles at the generator level before any
interaction with the detector. This can be related to the total cross
section, $\sigma$, through $\sigma_\text{fid} =  \sigma \, A$, where $A$ is the acceptance
in the fiducial volume.  Stable
particles are characterized as particles with mean lifetimes exceeding  30\unit{ps}. The upper limit on
$\sigma_\text{fid}$ is obtained from the limit on $\sigma \, A \, \epsilon$, where
$\epsilon$ accounts for detector resolution, trigger efficiencies, and identification and isolation requirements
applied in the analysis.

The leptons at the particle level are the electrons or muons
originating from the decay of W bosons. The charged leptons
from hadron decays are discarded, while electrons or muons from
direct decays of $\tau$ leptons  are included.

Stable particles, except muons, electrons, photons and neutrinos, are used to reconstruct
particle-level jets in the simulation. Jet reconstruction at the particle level is based on
the anti-\kt algorithm~\cite{kt}  with a distance parameter of 0.5. When a reconstructed jet contains a B hadron, the jet is
tagged as a b jet.
In events without a matched b jet, the jet with the largest \pt is used
to reconstruct the decayed top quark.
The \pt of the neutrinos is calculated as the magnitude of the vector sum of
the \pt of each neutrino in the event, except those originating from hadron decays.
From these objects, the top quark mass is calculated in order to make kinematical cuts
used in the definition of the fiducial region.
The fiducial region is introduced at particle level, similar to the
event selection requirements, and is summarized in Table \ref{fiducial}.

\begin{table*}[htb]
\centering
\topcaption{Definition of the fiducial region.
\label{fiducial}}
\begin{tabular}{ccccc}
\hline
     Object                 &  Requirement         \\
\hline
  Single muon          &  $\pt > 26$\GeV, $\abs{\eta}<2.1$        \\
  Veto for additional muons            &  $\pt > 10$\GeV, $\abs{\eta}<2.5$        \\
 Electron veto      &    $\pt > 20$\GeV, $\abs{\eta}<2.5$        \\
 Single photon                 &   $\pt > 50$\GeV,
 $\abs{\eta}<2.5$ ($1.44<\abs{\eta}<1.56$ excluded)       \\
 At least one jet ($N_\text{\PQb jet}<2$)                  &   $\pt > 30$\GeV, $\abs{\eta}<2.5$        \\
 Missing \pt                          & $\pt^\text{miss} > 30$\GeV \\
Muon, jets, and photons  &   $\Delta
R(\mu,\gamma)$ and  $\Delta
R(\text{jet},\gamma)>0.7$  \\
Reconstructed top quark mass  &  $130 < m_{\mu\nu \PQb}< 220$\GeV \\
\hline
\end{tabular}
\end{table*}

The efficiency $\epsilon$ is found to be 16\% and 19\% from simulation for the respective
$\PQt\PQu\gamma$ and $\PQt\PQc\gamma$ events in the fiducial region. An additional fiducial
region is defined by also requiring exactly one b-tagged jet in the event.
The values of $\epsilon$ are thereby reduced to 11\% and 14\%
for the two signals, respectively.

Table~\ref{fid} shows the 95\% CL upper limits on the signal cross
sections in the two fiducial regions for the $\PQt\PQu\gamma$ and $\PQt\PQc\gamma$
processes. These are calculated from the total number of selected
events in data ($N_\text{obs}$), the SM expectation ($N_\mathrm{SM}$),
both at detector level, and the efficiency for a signal event in the
fiducial region to be reconstructed at detector level.
The uncertainties in the SM expectation include
statistical and systematic uncertainties.
The total number of observed events is decreased by a factor of
approximately 6.5 after requiring exactly one identified b jet in an
event, while the expected number of SM events decreases by a factor of
7. The combined relative uncertainty in the number of expected SM
events increases from $12\%$ to $19\%$ when this b jet requirement is included.

The upper limits are calculated including a total systematic uncertainty
in the signal selection efficiencies of 10\%, estimated using a
method similar to that described in Section \ref{syssec}.
These are the first limits set on the anomalous $\PQt\gamma$ production within a
restricted phase-space region.

\begin{table}[h]
\centering
\topcaption{The total number of observed
  selected events at detector level in the data ($N_\text{obs}$), the SM expectations ($N_\mathrm{SM}$), the efficiencies ($\epsilon$),
and the upper limits on the cross sections $\sigma_\text{fid}$ at the
95\% CL in the fiducial region for the two signal channels, without and with a requirement on the presence of a single accompanying b jet.
 \label{fid}}
\begin{tabular}{lccccc}
\hline
Fiducial region                                   & Channel    &
$N_\text{obs}$            & $N_\mathrm{SM}$                       &
$\epsilon$ & $\sigma_\text{fid}^{95\%}$ (fb) \\ \hline
\multirow{2}{*}{Basic selection (Table \ref{fiducial})}  & $\PQt\PQu\gamma$ & \multirow{2}{*}{1794} & \multirow{2}{*}{$1805\pm 215$} & $0.16$     & 122                \\ \cline{2-2} \cline{5-6}
                                                  & $\PQt\PQc\gamma$ &                       &                                & 0.19     & 103               \\
\multirow{2}{*}{Basic selection and $N_\text{\PQb jet}=1$} & $\PQt\PQu\gamma$ &
\multirow{2}{*}{275}  & \multirow{2}{*}{$258 \pm 49$} & 0.11     & 47                 \\ \cline{2-2} \cline{5-6}
                                                  & $\PQt\PQc\gamma$ &
                                                  &
                                                  &  $0.14$          &
                                                  39              \\  \hline
\end{tabular}

\end{table}

\section{Summary}

The result of a search for flavor changing neutral
currents (FCNC) through single top
quark production in association with a photon has been presented.
The search is performed using proton-proton collisions at
a center-of-mass energy of 8\TeV, corresponding to an integrated luminosity of 19.8\fbinv, collected by the CMS detector at
the LHC. The number of observed events is consistent with
the SM prediction.
Upper limits are set at 95\% CL on the anomalous FCNC couplings of
$\kappa_{\PQt\PQu\gamma} <0.025$ and $\kappa_{\PQt\PQc\gamma} < 0.091$
using NLO QCD calculations.
The corresponding upper limits on the branching fractions are $\mathcal{B}(\PQt\to\PQu\gamma)< 1.3\times 10^{-4}$
and $\mathcal{B}(\PQt\to\PQc\gamma) < 1.7\times 10^{-3}$, which are
the most restrictive bounds to date.
Observed upper limits on the cross section in a restricted phase space
are found to be 47\unit{fb} and 39\unit{fb} at 95\% CL for $\PQt\PQu\gamma$ and $\PQt\PQc\gamma$
production, respectively, when exactly one identified b jet is required in the data.
These are the first results on anomalous $\PQt\gamma$ production within a
restricted phase-space region.

\begin{acknowledgments}
We congratulate our colleagues in the CERN accelerator departments for the excellent performance of the LHC and thank the technical and administrative staffs at CERN and at other CMS institutes for their contributions to the success of the CMS effort. In addition, we gratefully acknowledge the computing centers and personnel of the Worldwide LHC Computing Grid for delivering so effectively the computing infrastructure essential to our analyses. Finally, we acknowledge the enduring support for the construction and operation of the LHC and the CMS detector provided by the following funding agencies: BMWFW and FWF (Austria); FNRS and FWO (Belgium); CNPq, CAPES, FAPERJ, and FAPESP (Brazil); MES (Bulgaria); CERN; CAS, MoST, and NSFC (China); COLCIENCIAS (Colombia); MSES and CSF (Croatia); RPF (Cyprus); MoER, ERC IUT and ERDF (Estonia); Academy of Finland, MEC, and HIP (Finland); CEA and CNRS/IN2P3 (France); BMBF, DFG, and HGF (Germany); GSRT (Greece); OTKA and NIH (Hungary); DAE and DST (India); IPM (Iran); SFI (Ireland); INFN (Italy); MSIP and NRF (Republic of Korea); LAS (Lithuania); MOE and UM (Malaysia); CINVESTAV, CONACYT, SEP, and UASLP-FAI (Mexico); MBIE (New Zealand); PAEC (Pakistan); MSHE and NSC (Poland); FCT (Portugal); JINR (Dubna); MON, RosAtom, RAS and RFBR (Russia); MESTD (Serbia); SEIDI and CPAN (Spain); Swiss Funding Agencies (Switzerland); MST (Taipei); ThEPCenter, IPST, STAR and NSTDA (Thailand); TUBITAK and TAEK (Turkey); NASU and SFFR (Ukraine); STFC (United Kingdom); DOE and NSF (USA).

Individuals have received support from the Marie-Curie program and the European Research Council and EPLANET (European Union); the Leventis Foundation; the A. P. Sloan Foundation; the Alexander von Humboldt Foundation; the Belgian Federal Science Policy Office; the Fonds pour la Formation \`a la Recherche dans l'Industrie et dans l'Agriculture (FRIA-Belgium); the Agentschap voor Innovatie door Wetenschap en Technologie (IWT-Belgium); the Ministry of Education, Youth and Sports (MEYS) of the Czech Republic; the Council of Science and Industrial Research, India; the HOMING PLUS program of the Foundation for Polish Science, cofinanced from European Union, Regional Development Fund; the OPUS program of the National Science Center (Poland); the Compagnia di San Paolo (Torino); MIUR project 20108T4XTM (Italy); the Thalis and Aristeia programs cofinanced by EU-ESF and the Greek NSRF; the National Priorities Research Program by Qatar National Research Fund; the Rachadapisek Sompot Fund for Postdoctoral Fellowship, Chulalongkorn University (Thailand); and the Welch Foundation, contract C-1845.
 \end{acknowledgments}

\bibliography{auto_generated}

\cleardoublepage \appendix\section{The CMS Collaboration \label{app:collab}}\begin{sloppypar}\hyphenpenalty=5000\widowpenalty=500\clubpenalty=5000\input{TOP-14-003-authorlist.tex}\end{sloppypar}
\end{document}

%% file: TOP-14-003-authorlist.tex
\textbf{Yerevan Physics Institute,  Yerevan,  Armenia}\\*[0pt]
V.~Khachatryan, A.M.~Sirunyan, A.~Tumasyan
\vskip\cmsinstskip
\textbf{Institut f\"{u}r Hochenergiephysik der OeAW,  Wien,  Austria}\\*[0pt]
W.~Adam, E.~Asilar, T.~Bergauer, J.~Brandstetter, E.~Brondolin, M.~Dragicevic, J.~Er\"{o}, M.~Flechl, M.~Friedl, R.~Fr\"{u}hwirth\cmsAuthorMark{1}, V.M.~Ghete, C.~Hartl, N.~H\"{o}rmann, J.~Hrubec, M.~Jeitler\cmsAuthorMark{1}, V.~Kn\"{u}nz, A.~K\"{o}nig, M.~Krammer\cmsAuthorMark{1}, I.~Kr\"{a}tschmer, D.~Liko, T.~Matsushita, I.~Mikulec, D.~Rabady\cmsAuthorMark{2}, B.~Rahbaran, H.~Rohringer, J.~Schieck\cmsAuthorMark{1}, R.~Sch\"{o}fbeck, J.~Strauss, W.~Treberer-Treberspurg, W.~Waltenberger, C.-E.~Wulz\cmsAuthorMark{1}
% ...
\vskip\cmsinstskip
% ...
1:~~Also at Vienna University of Technology, Vienna, Austria\\
2:~~Also at CERN, European Organization for Nuclear Research, Geneva, Switzerland\\
% ...
